\documentclass[manuscript]{acmart}
\AtBeginDocument{%
  }

\setcopyright{acmlicensed}
\copyrightyear{2025}
\acmYear{2025}
\acmDOI{XXXXXXX.XXXXXXX}

\acmConference[Conference acronym 'XX]{Make sure to enter the correct
  conference title from your rights confirmation emai}{March 24--27, 2025}{Cagliari, Italy}
\acmBooktitle{}
\acmISBN{}

\acmSubmissionID{4396}

\usepackage{kotex}
\usepackage{subfig}
\usepackage{multirow}

\begin{document}

\title{A Room to Roam: Reset Prediction Based on Physical Object Placement for Redirected Walking}


\author{Sulim Chun}
\email{slchun@yonsei.ac.kr}
\affiliation{%
  \institution{Yonsei University}
  \city{Seoul}
  \country{Republic of Korea}
}

\author{Ho Jung Lee}
\email{dearshwan@yonsei.ac.kr}
\affiliation{%
  \institution{Yonsei University}
  \city{Seoul}
  \country{Republic of Korea}
}

\author{In-Kwon Lee}
\authornote{Corresponding Author}
\email{iklee@yonsei.ac.kr}
\affiliation{%
  \institution{Yonsei University}
  \city{Seoul}
  \country{Republic of Korea}
}


\begin{abstract}
   In Redirected Walking (RDW), resets are an overt method that explicitly interrupts users, and they should be avoided to provide a quality user experience. The number of resets depends on the configuration of the physical environment; thus, inappropriate object placement can lead to frequent resets, causing motion sickness and degrading presence. However, estimating the number of resets based on the physical layout is challenging. It is difficult to measure reset frequency with real users repeatedly testing different layouts, and virtual simulations offer limited real-time verification. As a result, while rearranging objects can reduce resets, users have not been able to fully take advantage of this opportunity, highlighting the need for rapid assessment of object placement. To address this, in Study 1, we collected simulation data and analyzed the average number of resets for various object placements. In study 2, we developed a system that allows users to evaluate reset frequency using a real-time placement interface powered by the first learning-based reset prediction model. Our model predicts resets from top-down views of the physical space, leveraging a Vision Transformer architecture. The model achieved a root mean square error (RMSE) of $23.88$. We visualized the model's attention scores using heatmaps to analyze the regions of focus during prediction. Through the interface, users can reorganize furniture while instantly observing the change in the predicted number of resets, thus improving their interior for a better RDW experience with fewer resets.
   
\end{abstract}


\begin{teaserfigure}
  \includegraphics[width=\textwidth]{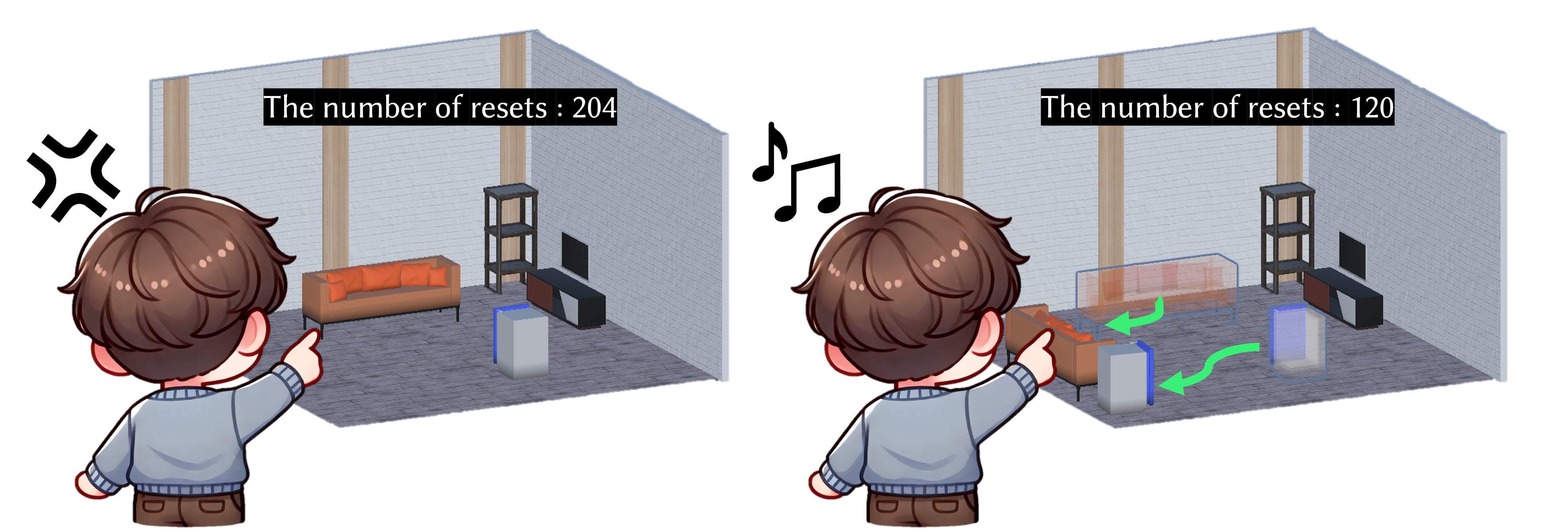}
  \caption{Illustration of the predicted number of resets in real time as the user adjusts the placement of object within our interface. The model we propose makes real-time predictions based on a top-down view image of the physical space. As shown on the left, when the object is positioned at the center of the physical space, the model predicts a high number of resets during RDW, indicating inefficient placement. On the right, the user, assisted by the system, arranges the objects in a way that results in fewer resets in their specific context.}
  \Description{Teaser}
  \label{fig:teaser}
\end{teaserfigure}

\received{11 October 2024}

\maketitle
\section{Introduction}

Virtual reality (VR) allows users to engage in immersive, three-dimensional environments through enhanced modalities, enabling extensive exploration of innovative interface systems. The locomotion interface is one of the critical aspects of VR that manages how users move and navigate in virtual space. As VR environments increase in complexity and scale, research has focused on finding the appropriate locomotion interface to enhance the user experience for various tasks \cite{di2021locomotion, kim2024locomotion, martinez2022research, martinez2022research}. Among various locomotion techniques, natural walking significantly enhances user immersion and improves navigation performance because it involves actual user movements \cite{usoh1999walking, ruddle2009benefits}. However, since many VR setups operate in confined physical spaces, natural walking also introduces the challenge of accommodating physical space constraints. To address the issue, Razzaque \cite{razzaquerdw} proposed Redirected Walking (RDW), which maps users' physical movements onto virtual spaces, allowing users to explore larger virtual environments without taking up as much physical space. This is achieved through subtle manipulations of the Head Mounted Display (HMD) screen or configuration controls of the virtual environment. When these subtle adjustments no longer avoid collisions, the user must overtly stop and reorient. A typical method is the reset technique, which is important for exploring large virtual spaces using RDW with small physical spaces \cite{williams2007exploring}. In many ways, RDW has been developed as a combination of subtle manipulation and reset techniques. However, the overt nature of resets can interrupt the user's presence \cite{suma2012taxonomy}. Therefore, optimizing RDW techniques to minimize the number of resets is a critical area of ongoing research.

According to previous studies, the number of resets varies with the size and shape of the physical space \cite{azmandian2015physical, messinger2019effects}. The physical space where a real user experiences VR through RDW is the environment (e.g., a room at home, a laboratory) with various obstacles such as different furniture. The rooms in a home are usually not large enough for VR experiences, and the complexity may increase with the placement of many pieces of furniture \cite{shin2019any, shin2022effects}. Therefore, the number of resets during the experience varies greatly depending on the placement of the object. The variability in reset frequency indicates that users have the potential to optimize their physical environment in order to reduce resets and enhance their overall experience. However, it is often not possible to optimize the physical space so effectively. This is largely because estimating the number of resets based on the configuration of the physical space typically relies on empirical methods, such as simulations or user studies, which are both time-consuming and require significant physical effort. Previous studies, such as the Environment Navigation Incompatibility (ENI) \cite{williams2022eni}, have focused on measuring navigability performance to improve user experience. However, there remains a lack of real-time methods and interfaces that can accurately predict resets and dynamically assist users in optimizing their physical space configuration.

In response, we propose an interactive system that allows users to experiment with object placement in real time, helping them to minimize 
resets. As shown in Fig.~\ref{fig:teaser}, the system allows users to drag and reposition furniture while receiving immediate feedback on the expected number of resets for each configuration. We developed a Vision Transformer (ViT) \cite{dosovitskiy2020image} based model to predict the number of resets from a top-view image of a physical space represented as object polygons. In Study 1, we ran RDW simulations in an environment with randomly arranged furniture, such as rooms in a home, to measure the number of resets in each arrangement. We confirmed that the number of resets is statistically significantly different depending on the number of objects placed in the physical space. We also found that for the same number of obstacle placements, the distribution of resets was broad and statistically different for different numbers of objects. The data collected in Study 1 is used as a training set for the model. In Study 2, we trained our model on the collected data and verified its prediction performance. Our model showed a result of $23.88$ for the Root Mean Square Error (RMSE), which represents the error in prediction performance. Furthermore, we employed heat maps to visualize and analyze the regions of high attention in the predictions of our ViT-based trained model. We expect that our model allows users to experience more immersive VR content by changing the layout of the physical space to reduce the number of resets. We summarize our contributions as follows:

\begin{enumerate}
\item We propose a novel interface that allows users to explore the object placement variations to reduce the number of resets according to their unique environment, enabling more immersive natural walking in VR environments.
\item To the best of our knowledge, we present a first learning-based method for predicting the number of resets for a specific environment in real time without user studies or simulations. 
\end{enumerate}

\section{Related Work}

\subsection{Redirected Walking}
Redirected Walking (RDW) is a locomotion interface for VR that enables natural walking in constrained physical spaces. RDW has been one of the most actively researched areas since it was first proposed by Razzaque \cite{razzaquerdw}. While methods using joystick control and teleportation systems have been extensively studied for their efficiency, RDW has consistently demonstrated a superior user experience in various applications \cite{ruddle2009benefits, langbehn2018evaluation, kim2024locomotion}. RDW achieves natural walking in confined physical spaces by subtly altering the user's perceived motion through manipulations of the VR camera view, with redirection gains controlling the magnitude of these adjustments. Steinicke et al.~\cite{steinicke2009estimation} identified three primary types of gains: translation gain, rotation gain, and curvature gain. Translation gain controls the speed of the user's forward movement in the virtual environment; rotation gain adjusts the speed of the user's rotation; and curvature gain subtly controls the user's physical orientation. These gains should be kept within a certain range because if the manipulation of the HMD screen is obvious and the user is aware that redirection gains are being applied, they may experience motion sickness and a decreased sense of presence \cite{steinicke2009real}. Steinicke et al.~\cite{steinicke2009estimation} measured the detection threshold below which the user is not aware of the redirection gain. Since then, there have been many studies that estimate detection thresholds for different situations and gains, and these studies provide a reasonable range that can be used in studies using RDW \cite{steinicke2009real, grechkin2016revisiting, kruse2018can}.
 
However, despite the use of these subtle manipulations, users inevitably encounter the boundaries of the physical environment, requiring overt techniques to prevent collisions. The reset technique, first proposed by Williams et al.~\cite{williams2007exploring}, is one of the most common methods used in such cases. Resets involve reorienting users, typically by turning them in place as they approach the boundaries of the physical space, which requires the use of gains that exceed the detection threshold. A typical example is a 2:1 turn reset. 2:1 turn provides a rotation gain that exceeds the detection threshold when a user rotates 180 degrees in place in the physical space, resulting in a 360 degree rotation in the virtual space. After resetting, the user can continue the experience facing away from the boundary in real space and in the direction they previously walked in virtual space.
While necessary to maintain continuous walking and ensure user safety, it is important to minimize resets that require redirection above the detection threshold. Fewer resets during the experience will allow users to maintain immersion and provide a seamless user experience. As a result, reset frequency has become a key performance measure for RDW. To minimize resets, various methods that optimize reset direction have been actively explored \cite{thomas2019general, zhang2022adaptive, lee2024marr, li2024enabling}. The proposed reset techniques, unlike the 2:1 turn, help reduce the number of resets by resetting the user in a specific direction depending on where the reset is performed. Typical examples include Reset to Gradient (R2G) and Modified Reset to Center (MR2C) \cite{thomas2019general}. MR2C resets the user to the center of the physical space, or perpendicular to the perimeter if the center is not available. R2G, on the other hand, resets the user in the direction of the sum of the virtual push forces applied to the user. Other strategies such as out-of-place resetting \cite{zhang2022one}, which resets to optimal position for free movement, and resetting to non-collision situations as needed, such as Point of Interests (PoIs) \cite{xu2022making}, have also been proposed. 

A method that combines subtle and reset techniques and manages their application is called a redirection controller. A representative method is the Artifical Potential Function (APF), which assumes that there are virtual pushing forces from walls or obstacles in the physical space and redirects the user to the vector of the sum of those forces \cite{thomas2019general}. The APF has been optimized and extended to be applicable to various situations such as multi-user and irregular spaces \cite{dong2020dynamic, chen2024apf}. Williams et al.~\cite{williams2021arc} presented an Alignment-based Redirection Controller (ARC) that considers both physical and virtual space and redirects the user to match the proximity of objects in the physical space around the user with the proximity of objects in the virtual space. Alignment-based redirection controllers have also been improved in various ways to reduce the number of resets and improve performance \cite{williams2021redirected, wu2023novel}.

On the other hand, apart from the performance improvement of the redirection controllers, the performance of the same controller will be different depending on the physical and virtual space experienced \cite{azmandian2015physical, williams2022eni, lee2024redirection}. Previous studies have investigated the effect of the size and shape of the physical space on the performance of the redirection controller. Azmandian et al.~\cite{azmandian2015physical} found that increasing the size of the tracked space and employing square-shaped areas can enhance RDW performance. Messinger et al.~\cite{messinger2019effects} observed that larger and concave-shaped spaces contribute to reducing the number of resets in an APF-based redirection controller. In addition, studies have attempted to develop navigability metrics based on environment layouts \cite{williams2021arc, williams2022eni}. However, there is no metric that can fully account for the number of resets. Therefore, to measure the performance of RDW, libraries such as the Redirected Walking Toolkit \cite{azmandian2016redirected} and Open-RDW \cite{li2021openrdw} are used to implement environments and redirection controllers for simulation or user experimentation. Azmandian et al.~\cite{azmandian2022validating} validated that simulation is an empirically valid evaluation method for redirected walking. However, given that simulation is time-consuming and user studies require elaborate environmental setup, a model capable of predicting RDW performance based on the environment is needed.

\subsection{Learning-based approach}

The rapid advancement of machine learning and deep learning has led to their increasing application in VR research to address various challenges \cite{hirzle2023xr}. Recent work on cybersickness prediction in VR has actively leveraged machine learning models \cite{kundu2023vr}, as well as deep learning-based approaches, such as Convolutional Neural Networks (CNNs) \cite{jeong2019cybersickness}, Long Short-Term Memory (LSTM) \cite{kim2019deep}, and Transformers \cite{jeong2024precyse}, to predict sickness based on inputs like Electroencephalogram (EEG) and eye gaze data. Gestural interaction support is also widely explored, with applications in gesture recognition \cite{chen2021gestonhmd, mo2021gesture, al2022vision} and hand trajectory prediction \cite{gamage2021so}.

In the area of locomotion interfaces, deep learning is being explored to enhance the existing techniques and improve the user experience. For walking in place (WIP), Hanson et al.~\cite{hanson2019improving} used a CNN-based model to distinguish between walking and standing to improve WIP. Ke et al.~\cite{ke2021larger} trained a Support Vector Machine (SVM) classification model to control the speed according to the user's leg and foot gestures. For Unintended Positional Drift (UPD), Brument et al.~\cite{brument2021understanding} proposed machine learning-based based UPD model using a Gaussian Mixture Model (GMM). In the context of backward movements, Paik et al.~\cite{paik2021feel} developed a deep learning model using CNN and Bi-Long Short-Term Memory (BiLSTM) layers to predict forward and backward movement using sensor data.

RDW is one of the most popular applications of deep learning approaches in locomotion interfaces \cite{hirzle2023xr}. A number of reinforcement learning approaches use the user's spatial information to train more effective redirection techniques. For example, reinforcement learning-based approaches have been used to determine the optimal target for redirecting the user rather than relying on a fixed center \cite{lee2019real}, and to prescribe the appropriate gain to apply \cite{strauss2020steering}. 
It is also used to predict the future position of the user to improve the performance of the redirection controller \cite{jeon2024f, lee2024redirection}.

On the other hand, the representation of scene layout has been extensively studied in areas such as indoor scene synthesis \cite{patil2024advances}, which focuses on generating and modifying the spatial organization of objects within a given environment. Common approaches include images \cite{wang2018deep, wang2019planit}, 3D scans \cite{avetisyan2020scenecad}, and graphs \cite{patil2021layoutgmn}. In our work, we adopted a top-down view image approach to effectively capture the geometric layout, which is known to influence RDW performance \cite{williams2022eni}. To encode the image representation, data-driven method is widely used. Wang et al.~\cite{wang2018deep} trained a convolutional prior for object placement using a semantic representation (depth, category, orientation) in a top-down view. Wang et al.~\cite{wang2019planit} also proposed a placement model that uses both graph relations and image-based spatial prior, using a CNN-based model.

While CNN-based approaches have been successful in learning spatial information, recent advances in transformers have demonstrated even stronger performance in vision tasks \cite{han2022survey}. The Vision Transformer (ViT), proposed by Dosovitskiy et al.~\cite{dosovitskiy2020image}, achieved remarkable performance in the field of image recognition by dividing the image directly into patches and training on the transformer architecture \cite{vaswani2017attention}. Recently, beyond image recognition, transformer-based models have been applied to tasks such as image segmentation \cite{ye2019cross} and object detection \cite{carion2020end}. Regarding scene layout, Paschalidou et al.~\cite{paschalidou2021atiss} proposed an autoregressive Transformer architecture for indoor layout synthesis. Building on these advances, we use ViT to model the complexity of multiple objects in physical space, capturing spatial relationships throughout the layout that affect RDW performance.


\section{Method}

\begin{figure}[t]
  \centering
  \includegraphics[width=0.8\textwidth]{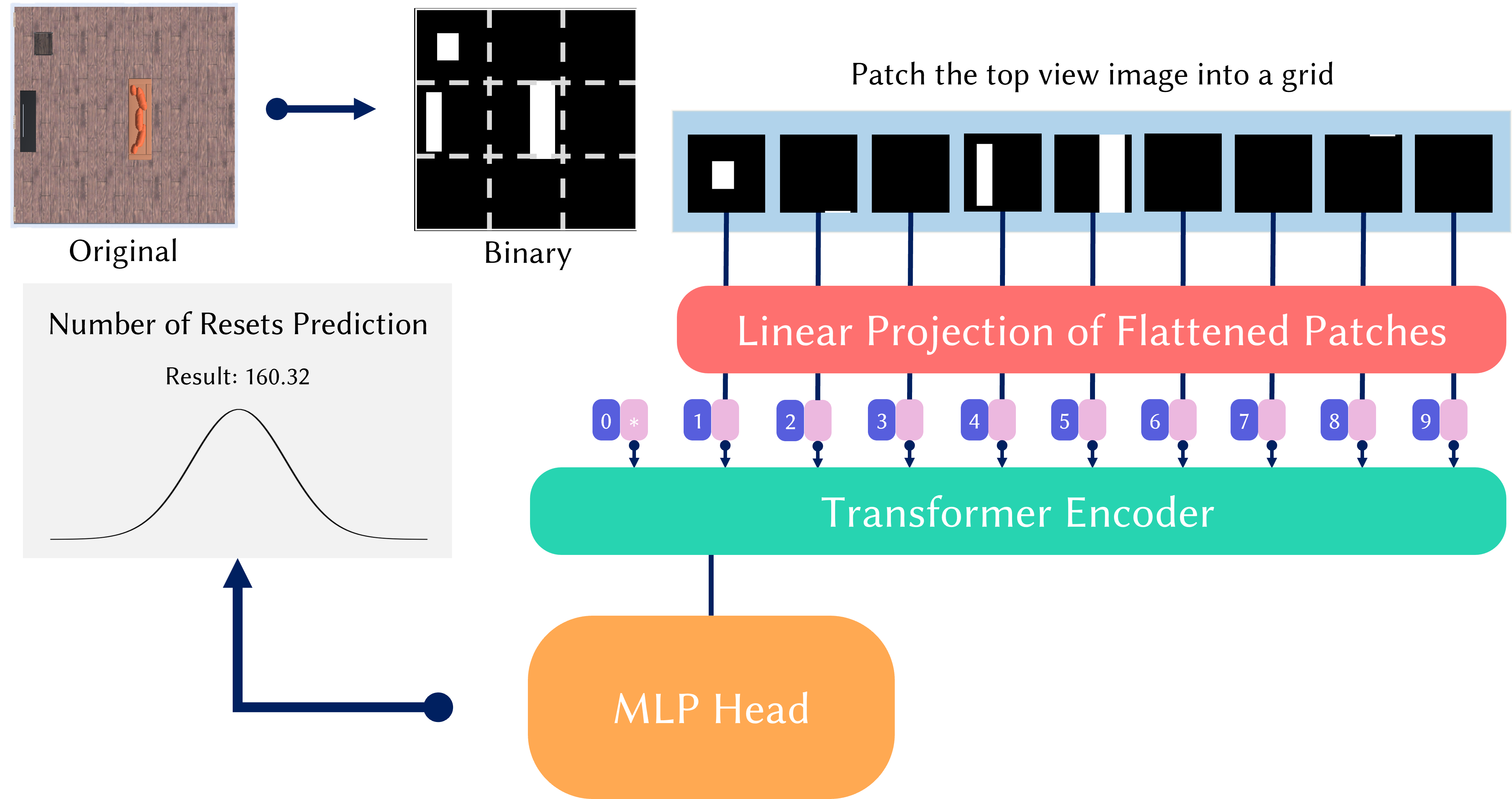}
  \caption{Overview of our reset prediction model. Our model predicts the number of resets an RDW user will experience in the given physical space based on the top-down view. The information of objects placed in the physical space is first transformed into a binary 
  top-down image, filled with 1s where objects are present and 0s where they are not. The image is fed into the Vision Transformer (ViT) model, where it is divided into patches of size 16 $\times$ 16 and embedded into a 768-dimensional space using linear projection. The image passes through 12 encoder layers of the transformer, and the number of resets is predicted through the Multi-Layer Perceptron (MLP) head. The data used to train the model was collected through simulation.}
  \label{fig:2overview}
  \Description{System Overview}
\end{figure} 

We developed a model to predict the number of resets the user will experience during RDW, based on the placement of objects in the physical space. A top-down view of a physical space was used to represent the placement of objects, accurately reflecting the boundaries of objects and walls, which are critical for observing the environment in RDW studies \cite{lee2019real, lee2020optimal}. Top-down view images have also proven useful in many deep learning-based methods for tasks such as indoor scene synthesis \cite{wang2018deep, wang2019planit} and robot navigation \cite{sharma2024pre}. In order to effectively capture the spatial context among multiple objects in the image and learn the corresponding resets, a robust deep learning model was required. We employed the Vision Transformer (ViT), which divides the image into patches and learns the relationships between them using a Transformer architecture. Vit has less image-specific inductive bias than CNN-based models as it does not employ any convolutional layers with locality and translation equivariance. Instead, it uses global self-attention layers followed by a multi-layer perceptron (MLP) head. This design allows the model to effectively learn the global features of our environment, where objects are distributed in space and each placement affects the number of resets. Among the model variants, 
we used the ViT-Base model with 16 $\times$ 16 input patch size (ViT-B/16) \cite{dosovitskiy2020image}.

Fig.~\ref{fig:2overview} shows the process of our system. First, to generate a top-down view image of the physical space where the objects are placed, the 2D polygon coordinates of the objects are scaled to size of 224 $\times$ 224 and converted to a binary image,
with a value of 1 for pixels containing objects and 0 for empty pixels. Since the ViT requires a three-channel input, we duplicated the single-channel binary image across all three channel without modifying the architecture. The ViT then divides the converted image into patches size of 16 $\times$ 16, and a linear projection is applied to create tokens with an embedding dimension of 768. After the positional embeddings are added, the tokens are passed through the 12 encoder layers, and are fed into the MLP head to predict the number of resets. Pre-trained weights from ImageNet-21k \cite{deng2009imagenet, kolesnikov2020big, ridnik2021imagenet} were used, as research suggests that using a pre-trained model rather than training from scratch is more effective and produces a better result \cite{dosovitskiy2020image, steiner2021train}.

To fine-tune a pre-trained ViT for our task, a sufficient amount of data is required. However, collecting such a large amount of data through real user experience while modifying the physical configuration of spaces is both time-consuming and impractical. Therefore, we used a more efficient simulation method that has been validated for RDW performance evaluation \cite{azmandian2022validating}. However, since our goal is to develop a system that can be used in real-world scenarios, we modeled the environment to resemble a typical room. Specifically, we collected simulation data to map the expected number of resets for the top-down view, which accurately represents the boundaries and positions of objects, and used the data for training. The following section (Section. ~\ref{sec:study1}) describes the data collection process and the results.

\section{Study 1}
\label{sec:study1}
\subsection{Data Collection}
\label{subsec:study1_data_collection}

\begin{figure}[ht!]
    \centering
    \captionsetup[subfloat]{}
    \subfloat[Top view\label{F3_1}]{\includegraphics[width=0.3\textwidth]{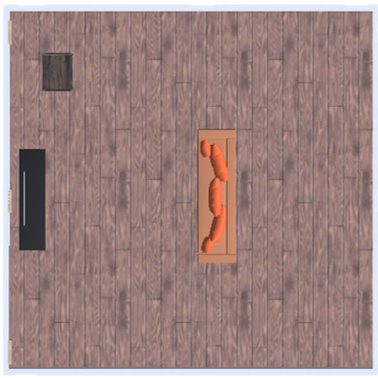}}
    \hspace{10mm}
    \subfloat[Side view\label{F3_2}]{\includegraphics[width=0.3\textwidth]{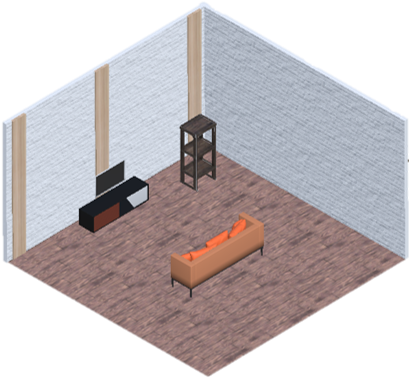}}
    \caption{The figure shows one of the simulation environments in which the data collection was carried out.}
    \label{fig:3realview}
    \Description{}
    \vspace{0mm}
\end{figure}

\begin{figure}[ht!]
    \centering
    \captionsetup[subfloat]{}
    \subfloat[Placed three objects in physical space.\label{F4_1}]{\includegraphics[width=0.45\textwidth]{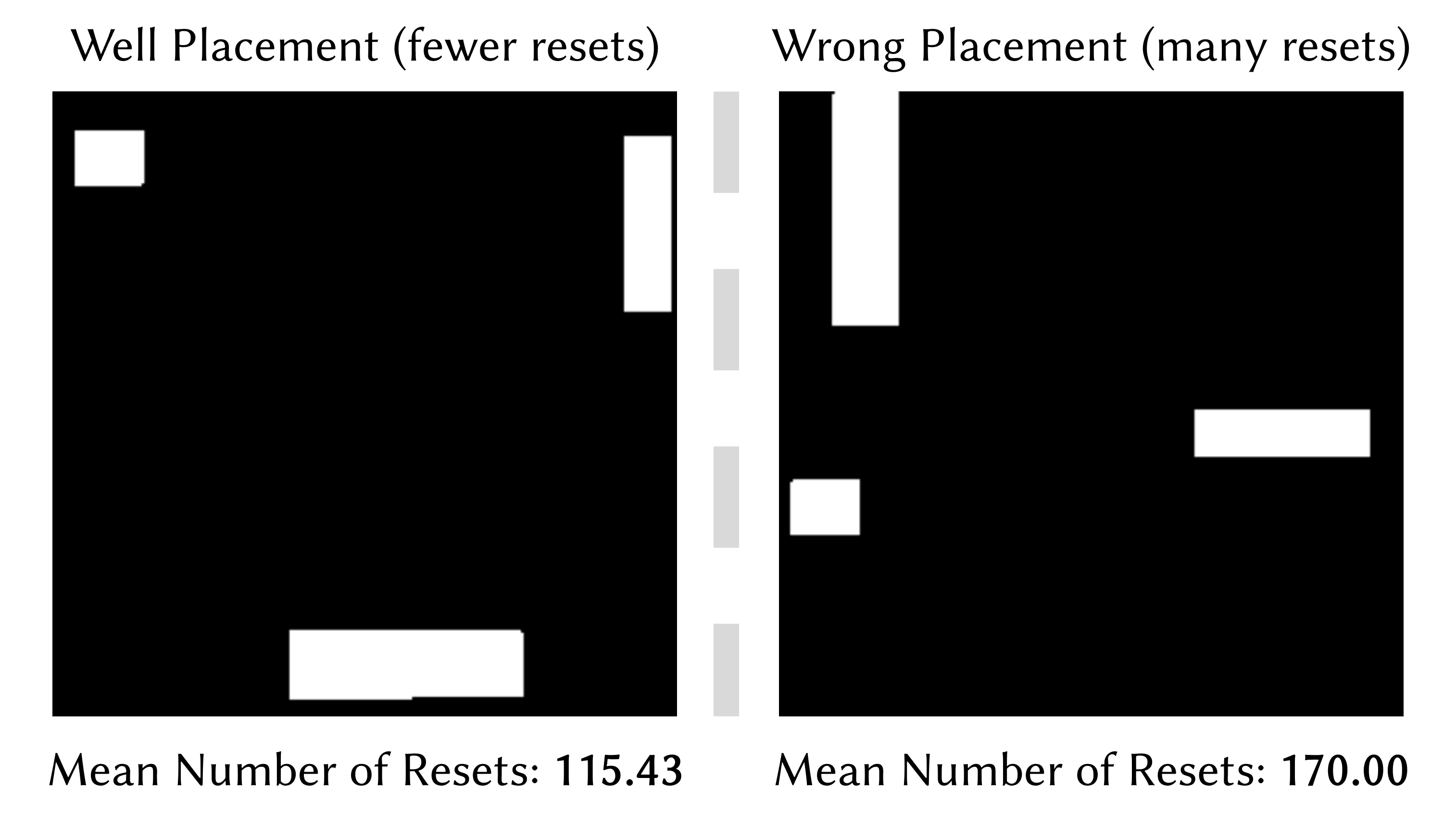}}
    \hspace{5mm}
    \subfloat[Placed five objects in physical space.\label{F4_3}]
    {\includegraphics[width=0.45\textwidth]{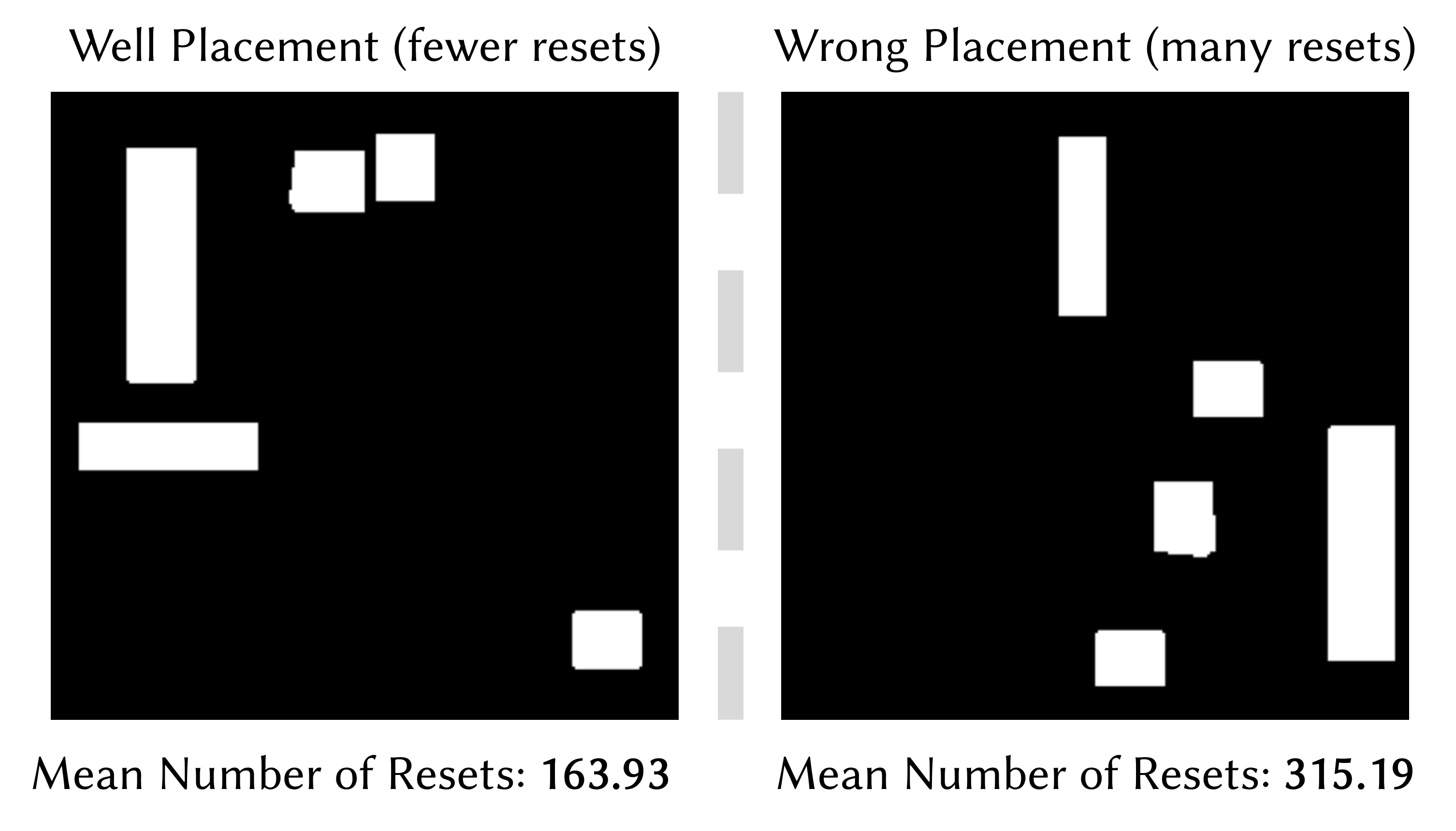}}
    \caption{An illustration where the number of resets changes depending on the placement of obstacles in the physical space. The top-down view displays objects in white and empty space in black. }
    \label{fig:4MReset}
    \Description{}
\end{figure}

In Study 1, we collected data through simulation to train our model. The physical space was modeled as a typical living room in a family home, considering the common indoor environment in which RDW is conducted. The size of the physical space was set to 5 $\times$ 5 meters, which is appropriate for conducting RDW without deviating from the dimensions of a standard living room \cite{fu20213d}. Five pieces of furniture of four different types were used to ensure that the top-down view resembled a common living room environment: a TV stand, a sofa, two shelves, and a mini-fridge \cite{fu20213d, shin2019any, shin2022effects, kim2023edge}. The furniture was either fully arranged or three to four randomly selected pieces were placed. Fig.~\ref{fig:3realview} shows an example of the arrangement of the physical space. The position of each piece of furniture was randomized within the boundaries of the room, and its rotation was randomized between 0, 90, 180, and 270 degrees.

Using the modeled environment, we developed an RDW simulation based on the Open-RDW library \cite{li2021openrdw}. The simulation was performed on an Intel i7-12700 CPU, NVIDIA RTX 4080 GPU and 32 GB of RAM, running at 90 frames per second. The translation and rotation speeds of the simulated user were 1.4$m/s$ and $90^{\circ}/s$, respectively. For the simulated user, we applied Thomas et al.'s APF (TAPF) and R2G \cite{thomas2019general}, which are commonly used redirection controllers in recent studies. The detection thresholds were set according to previous studies \cite{steinicke2009estimation, steinicke2009real, langbehn2017bending, williams2019estimation, lee2024marr}: $[$0.86, 1.26$]$ for translation gains, $[$0.67, 1.24$]$ for rotation gains, and 7.5 meters for curvature gains. In the simulation, the user starts at a random location in physical space and walks in a straight line to collect virtual targets, which are successively generated in a radius of 2-6 meters around each collected point. The simulation ends when the user has walked a total of 500 meters. The virtual space traversed by the user was set to be continuously expansive and empty, since the TAPF and R2G do not take into account virtual space information. Since the number of resets depends on the path taken by the user in the virtual environment, we aimed to collect enough samples with a variety of potential virtual paths to obtain a representative reset count for each physical object placement. Therefore, we configured the data by averaging the number of resets by having the simulated user walk 30 random paths for each placement. In each placement, the 2D polygon coordinates of all objects were recorded along with the reset data to construct the model input. Fig.~\ref{fig:4MReset} shows examples of the data for different object placements and the corresponding number of resets.

\subsection{Result}

We collected a total of 300 data entries, with 100 for each placement configuration of three, four, and five objects. Using the dataset, we analyzed the effect of object placement on the number of resets. First, we conducted statistical tests to determine whether the number of resets significantly with the number of objects. We used the Shapiro-Wilk and Levene's tests to assess normality and homogeneity of variance, but neither assumption was met, so we performed the non-parametric Kruskal-Wallis test. We found that the number of objects had a significant effect on the number of resets ($\chi^2 = 93.66,  p < .001, \eta^2 = 0.31$). A post hoc test using the Dunn test was also performed to confirm the differences between each number of objects. Bonferroni correction was performed to ensure statistical robustness. The results showed that all pairs (three-four, three-five, and four-five objects) were significantly different ($p < .001$). Thus, the data confirmed that changing the number of placed objects made a significant difference in the number of resets. Furthermore, Levene's test confirmed that the variances were not equal across the different numbers of objects ($\chi^2 = 29.56, p < .001, \eta^2 = 0.17$). We performed a Tukey's HSD post hoc test to determine whether the variance of the number of resets differed across the number of objects. There was a statistically significant difference in the variance when three obstacles were placed, compared to four and five objects ($p < .001$). These results show that the distribution of the number of resets is different even when the same number of objects are placed, indicating that reset prediction in multiple environments is complex but essential.

\begin{figure}[t]
    \centering
  \includegraphics[width=0.6\textwidth]{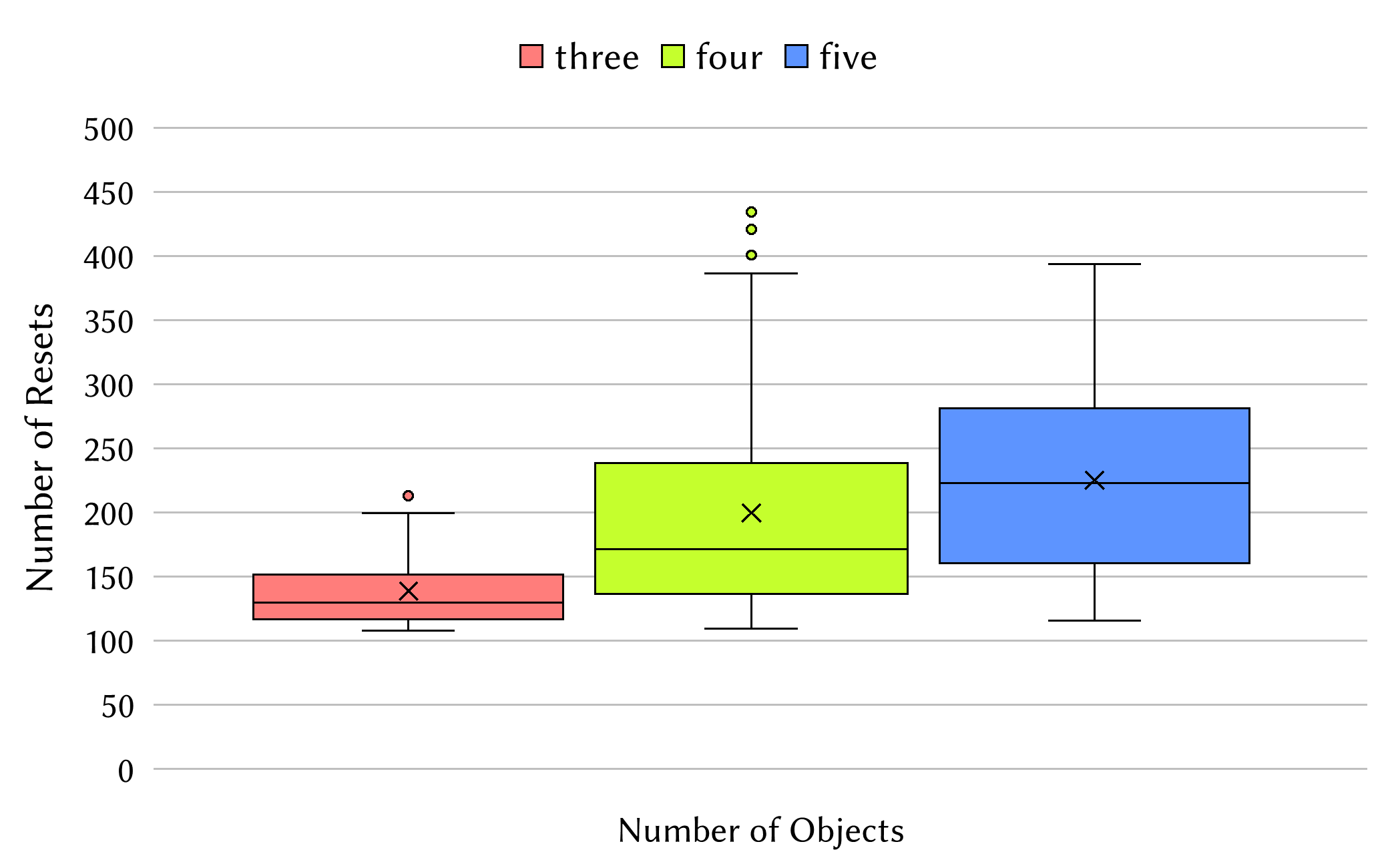}
  \caption{Mean and standard deviation ($SD$) results of the number of resets as a result of placing the same number of obstacles in physical space. Error bars indicate 95\% confidence intervals.}
  \label{fig:study1result}
  \Description{Result of study1}
\end{figure}

    


\section{Study 2}

\subsection{Training}

In Study 2, we trained a reset prediction model, evaluated its performance and developed an interactive user interface. The model was implemented using PyTorch \cite{NEURIPS2019_9015} version 2.1.0 with CUDA 11.8 (cu118) support and exported to ONNX for use in Unity. The interface was built using Unity version 2022.3.16f1, and the model inference was performed using the Barracuda package version 3.0.0. Both model training and interface development were conducted on an Intel i7-12700 CPU, NVIDIA RTX 4080 GPU and 32GB of RAM.

The model was trained using the dataset constructed in Section~\ref{sec:study1}. The entire dataset was used to train a single model, which is capable of processing physical space regardless of the number (3-5) and type of objects. The dataset was randomly split into training, validation, and test sets in a 6:2:2 ratio. The validation set was used to determine hyperparameters using a random grid search method, which has been shown to be efficient for hyperparameter optimization \cite{bergstra2012random}. The selected hyperparameters are detailed in Table~\ref{table:hyperparameter}. The test set was reserved for evaluation exclusively, and was not used at any stage of model training. For the learning rate, we applied the one-cycle learning rate policy \cite{smith2019super}, which increases the learning rate to a maximum and then anneals down. The maximum learning rate was set as specified in Table~\ref{table:hyperparameter}, and all other parameters for the were kept at the PyTorch default settings. An early stopping method was used to stop training if the validation loss did not decrease for 30 consecutive epochs. To improve the robustness of the model, we applied data augmentation techniques, including horizontal flipping, vertical flipping, and rotations of 90, 180, and 270 degrees, each with a probability of 5\%. 


\begin{table}[t]
\caption{Hyperparameter table}
\renewcommand{\arraystretch}{1.35}
\resizebox*{0.5\textwidth}{!}{%
\begin{tabular}{|c|c|}
\hline
\textbf{Hyperparameter}                & \textbf{Value}                                                          \\ \hline \hline

Batch size                    & 64                                                             \\ \hline
Learning rate                 & $1.0 \times 10^{-6}$                                              \\ \hline
Epoch                         & 500                                                            \\ \hline
Optimizer                     & Stochastic Gradient Descent (SGD)                              \\ \hline
Weight decay                  & $1.0 \times 10^{-4}$                                              \\ \hline
\end{tabular}
}
\label{table:hyperparameter}
\Description{}
\end{table}

\subsection{Result}

We evaluated the reset prediction model using Root Mean Squared Error (RMSE), Mean Absolute Error (MAE), and $R^2$ score, which are commonly used metrics to evaluate the performance of regression models. RMSE indicates the average size of prediction errors, with larger errors being penalized, while MAE provides the average of absolute errors, providing an overall measure of the deviation of predictions from actuals. The $R^2$ value is the proportion of variance that can be explained by the model, with values close to 1 representing better prediction performance.

\begin{figure}[ht!]
    \centering
    \captionsetup[subfloat]{}
    \subfloat[Validation\label{F6_1}]{\includegraphics[width=0.35\textwidth]{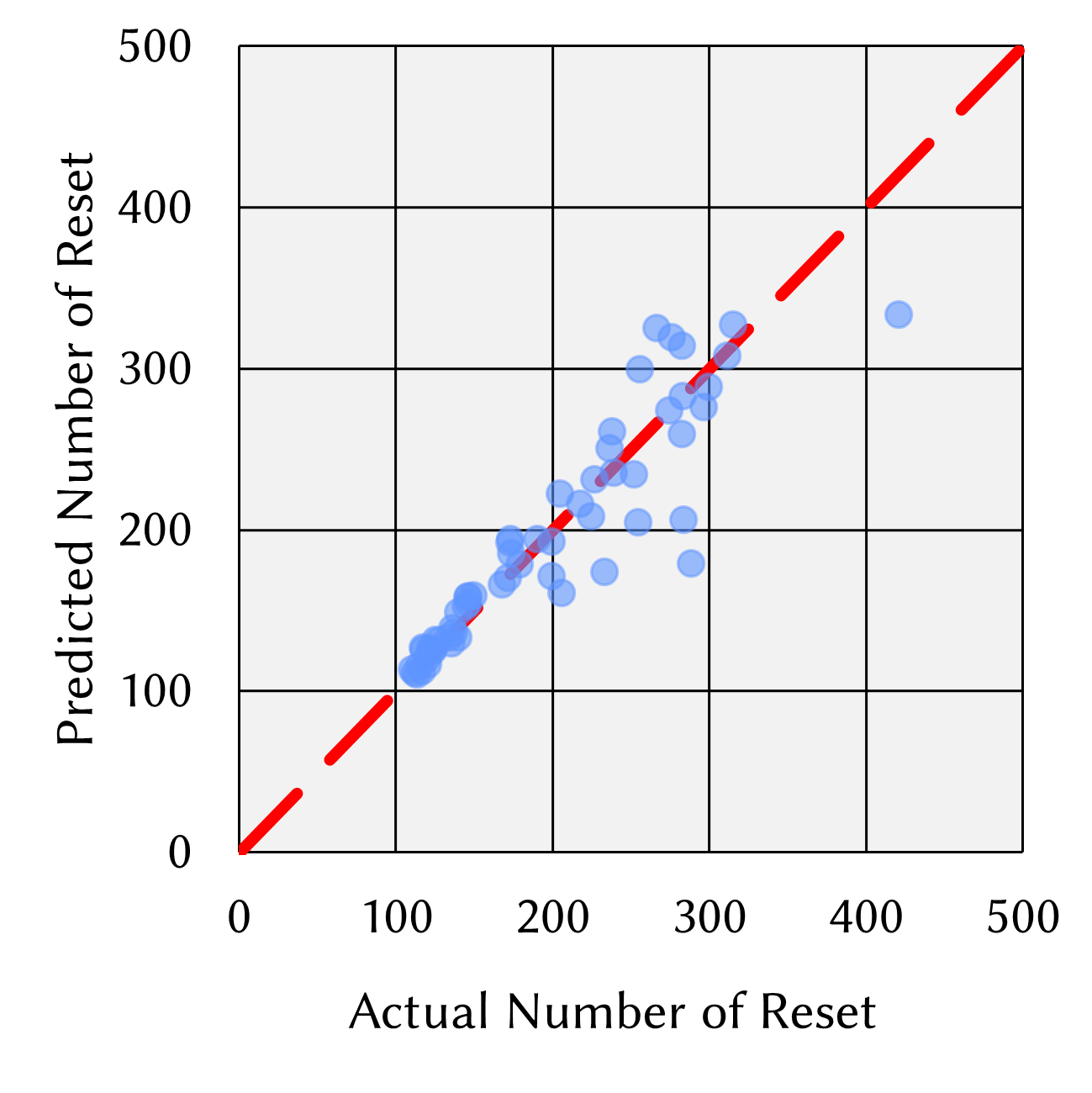}}
    \hspace{5mm}
    \subfloat[Test\label{F6_2}]{\includegraphics[width=0.35\textwidth]{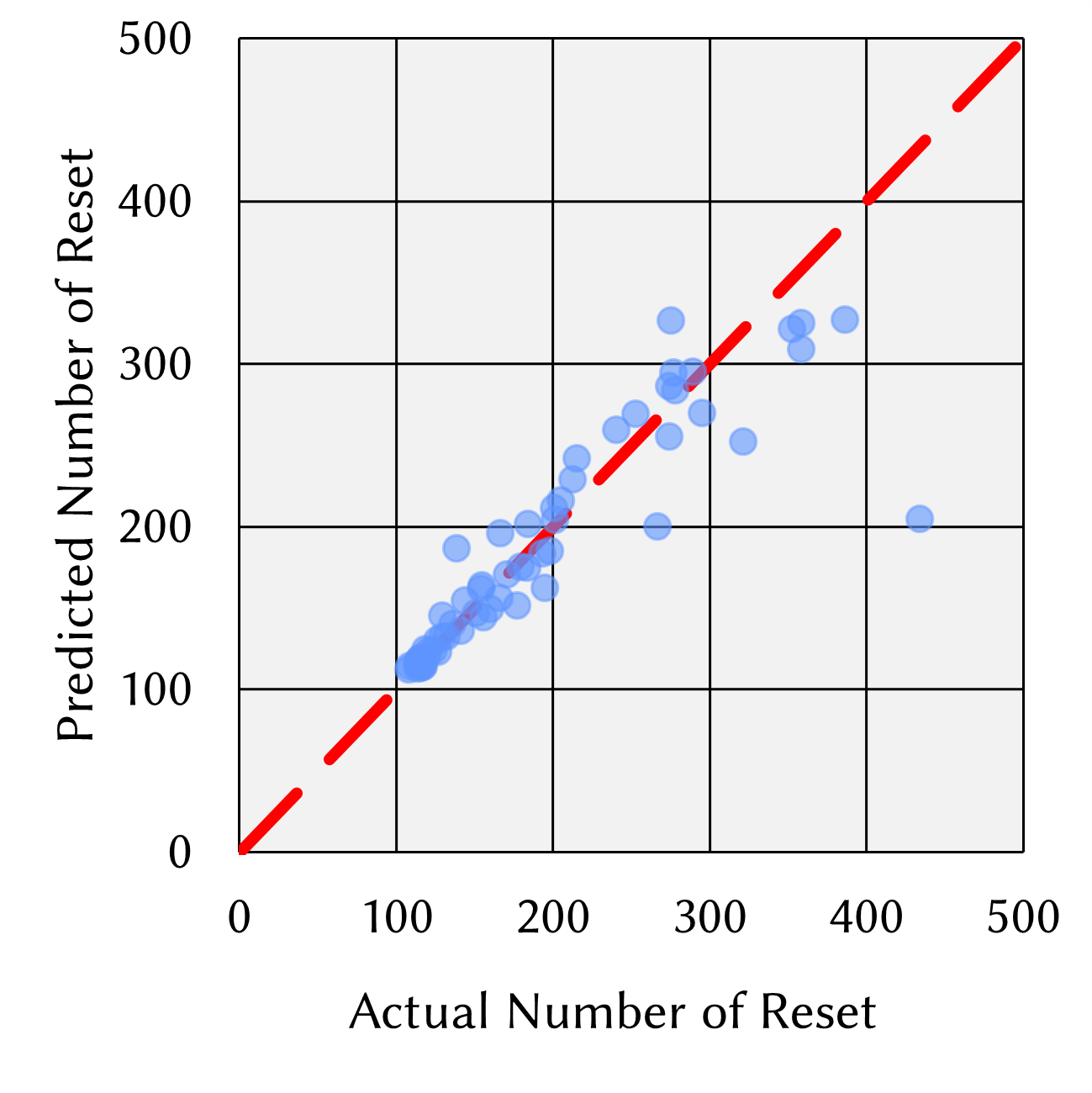}}

    \caption{Scatter plot illustrating the performance of the trained model on both the validation and test sets. The blue points represent the combination of the actual number of resets ($x$-axis) and the predicted number of resets ($y$-axis) for each data point. The red line indicates where the predicted values would be exactly the same as the actual values.}
    \label{fig:6performance}
    \Description{}
    \vspace{0mm}
\end{figure}

The evaluation of our model on the test dataset resulted in an RMSE of 23.88, an MAE of 15.36, and a $R^2$ score of 0.91. The predicted number of resets and the actual number of resets for all validation and test sets are plotted in Fig.~\ref{fig:6performance}. The x-axis represents the actual number of resets and the y-axis represents the predicted number of resets, with the actual values on the x-axis and the predicted values on the y-axis. The blue points represent individual data points, and the red line indicates where the predicted and actual values are equal.

Using our model, we implemented a system that allows the user to adjust the position of objects in a Unity environment and observe the predicted number of resets through the user interface. Users can move or rotate objects by dragging them, while the system generates a top-down image based on the updated object placement, predicts the number of resets using the model, and displays the result in the interface. This real-time feedback allows the user to interactively adjust the physical layout to account for resets.

\section{Discussion}
\subsection{Result Analysis}

We developed a reset prediction model using Vision Transformer (ViT) with low error rates, and proposed an interactive interface with real-time feedback to help users find the appropriate placement. Our data-driven approach provides a more time-efficient alternative to user studies or simulations, allowing users to estimate expected resets in their specific environments.

According to the results of Study 1, the number of resets varies significantly depending on the number of objects. Since a reset occurs when a user approaches an object or boundary in physical space, a denser environment with more physical objects can generally result in more resets. However, when looking at the respective distributions of the number of resets, there is a wide range even within the same number of objects. The difference in variance between 3-4 objects and 3-5 objects is also statistically significant. The user can have all five objects and still get a similar benefit as having three, if placed well. We trained a single model to handle all the data with different numbers of objects placed, showing the flexibility of our data-driven approach in to handle different scenarios. It should be noted, however, that this reset behaviour may change depending on the configuration of the virtual environment. Williams et al.~\cite{williams2022eni} proposed ENI, a metric that measures navigation compatibility based on the similarity between virtual and physical spaces using visibility polygons. This study showed that when the complex physical space remains fixed, compatibility can increase as the virtual space becomes more complex and similar to the physical space. 

According to the results of Study 2, our model appears to closely adhere to the regression trend, showing low error on the test dataset. However, Fig.~\ref{fig:6performance} shows that some data points with relatively high reset counts resulted in large errors. These higher reset counts are often caused by the user getting stuck in a tight space between obstacles in the physical space, resulting in repeated resets in some samples. Due to the limited number of such data (Fig.~\ref{fig:study1result}), the model struggled to fit this portion of the data accurately.

\begin{figure}[t]
    \centering
    \captionsetup[subfloat]{}
    \subfloat[\label{F7_1}]{\includegraphics[width=0.48\textwidth]{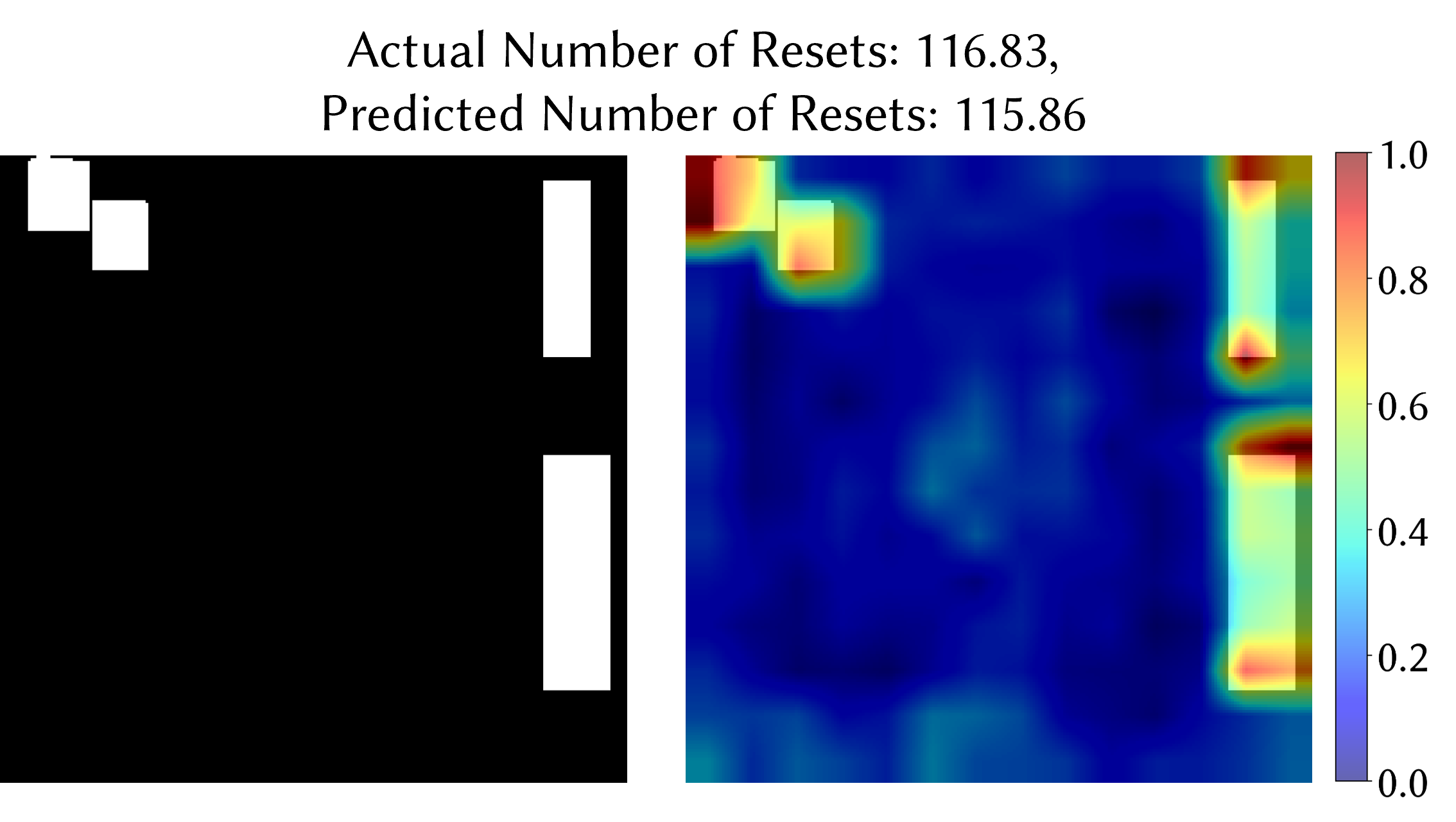}}
    \hfill
    \subfloat[\label{F7_2}]{\includegraphics[width=0.48\textwidth]{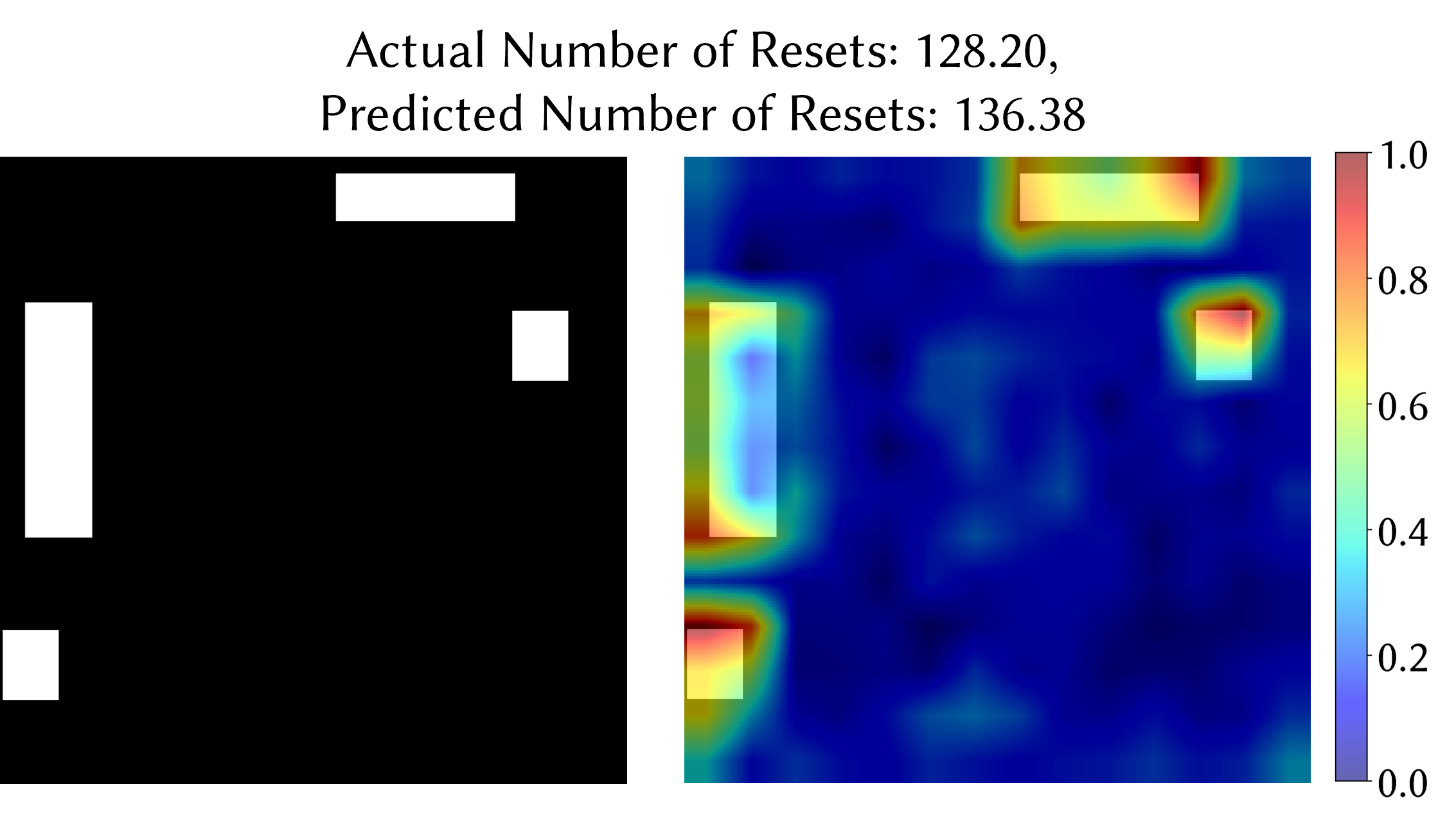}}

    \subfloat[\label{F7_3}]{\includegraphics[width=0.48\textwidth]{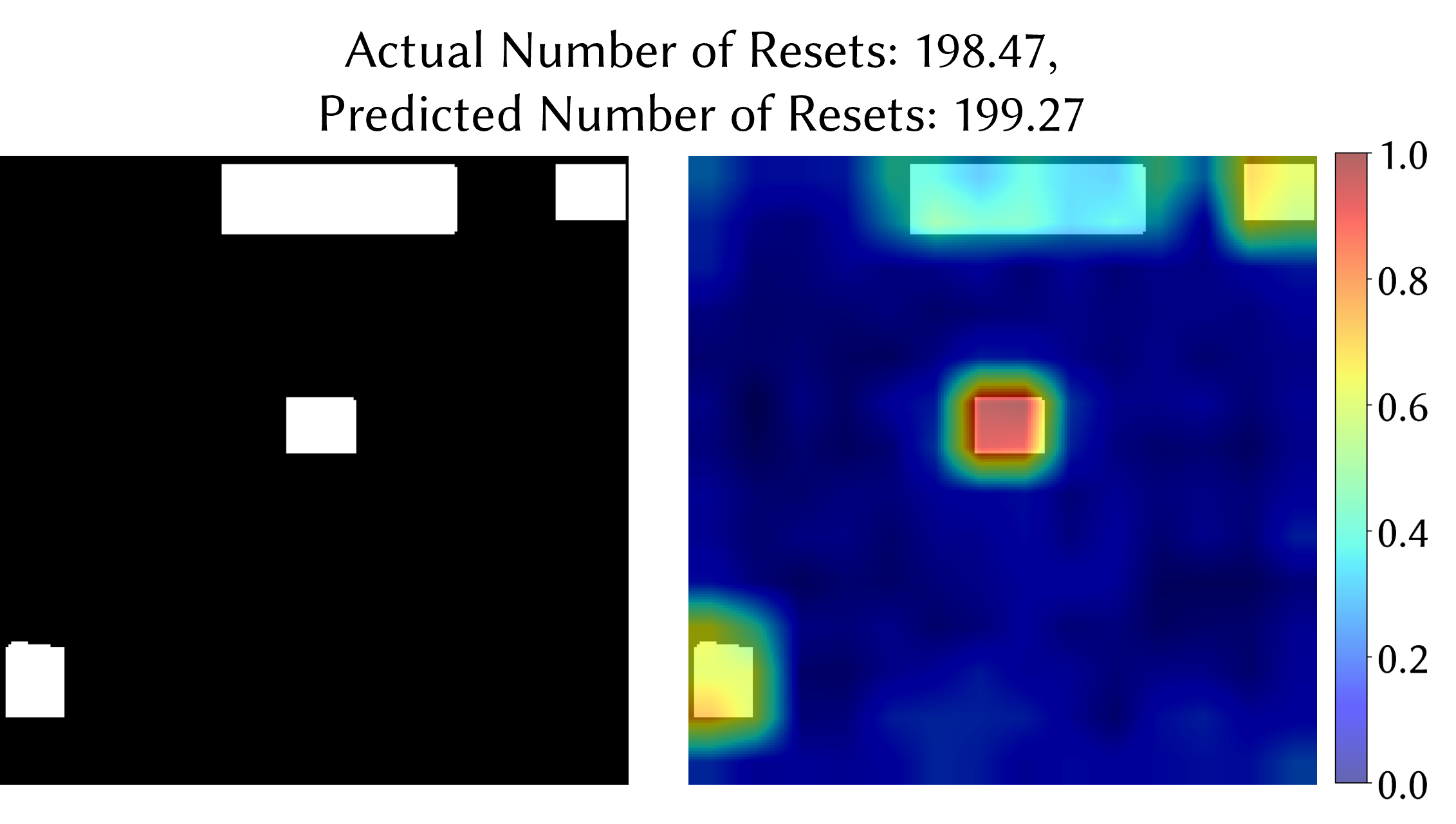}}
    \hfill
    \subfloat[\label{F7_4}]{\includegraphics[width=0.48\textwidth]{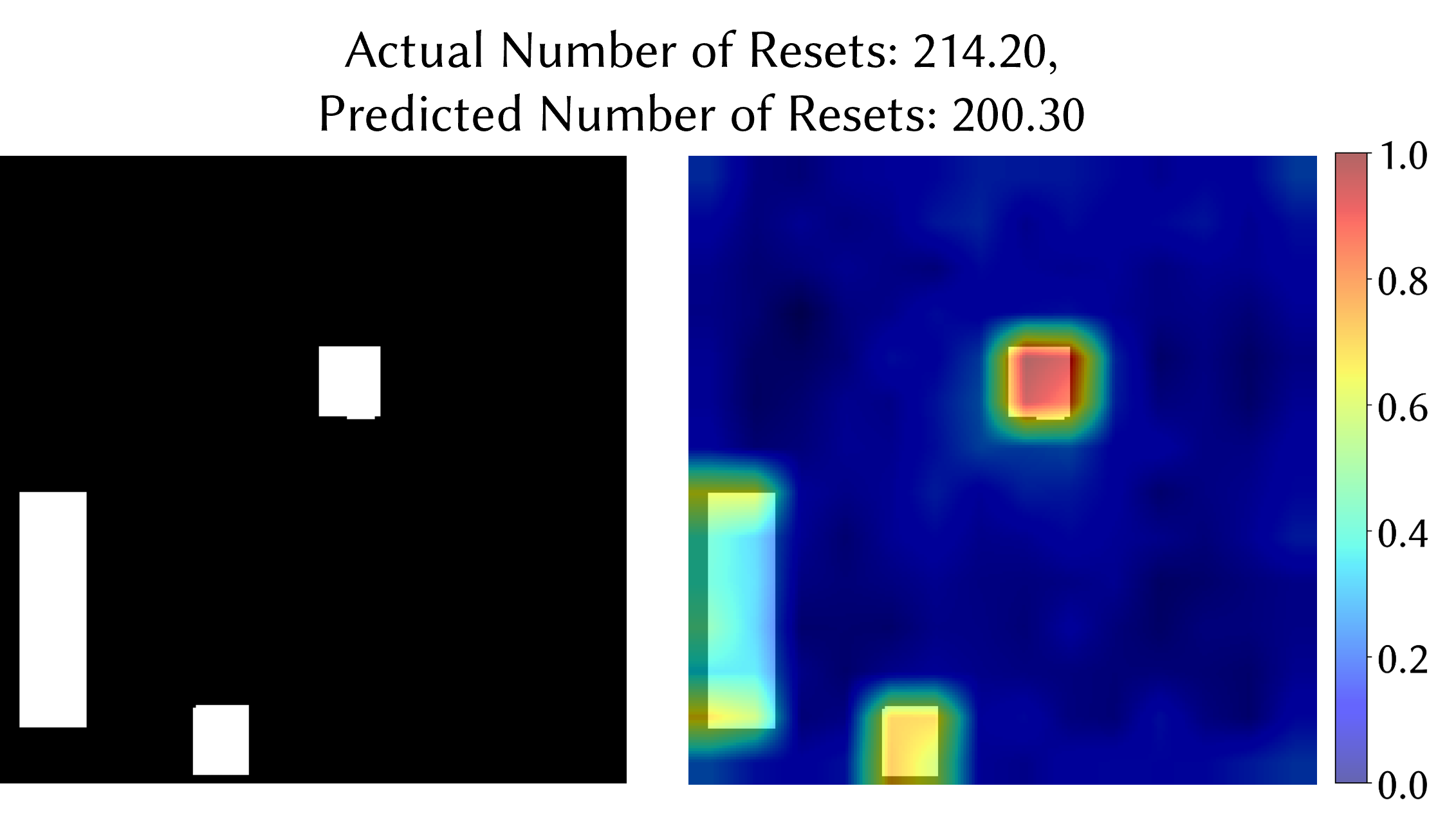}}
    \caption{Visualizations of attention weights. The left side of each subfigure shows a top-down view of the object placement, where white represents the objects and black represents empty space. On the right is a heatmap overlay of the attention weights calculated using the rollout method, with red areas indicating the highest attention and blue areas indicating the lowest attention. }
    \label{fig:7discussion}
    \Description{}
    \vspace{0mm}
\end{figure}

To examine the model's decisions, we visualized the attention weights from the ViT model \cite{vaswani2017attention, wiegreffe2019attention, vashishth2019attention, chefer2021transformer}. We used the attention rollout method \cite{abnar2020quantifying}, which recursively multiplies the attention weight matrices and takes into account the residual connections. Fig.~\ref{fig:7discussion} shows the top-down view of each placement on the left, with the corresponding attention heatmap overlaid on the right. Fig.~\ref{F7_1} and Fig.~\ref{F7_2} both illustrate a scenario where the actual number of resets is relatively minimal in the data (Fig.~\ref{fig:study1result}). The model predicted resets well, focusing on the edges of objects and areas adjacent to the walls, which are highlighted in red due to higher attention weights. We could also see that the corners of the objects are highlighted more on the wall than in the open space. Both Fig.~\ref{F7_3} and Fig.~\ref{F7_4} have a high number of actual resets and predicted resets, where the object in the middle is highlighted. In contrast to Fig.~\ref{F7_1} and Fig.~\ref{F7_2}, the attention is less concentrated on the edges and areas adjacent to the wall.




\subsection{Limitation and Future Work}

\textbf{Diversity in physical space.} We collected the dataset based on the layout of a typical living room in a family home. However, actual users will encounter a wider variety of room shapes and furniture types. In addition, RDW can be performed in other environments, such as larger studios and research laboratories. While our model successfully predicted the number of resets in the living room setting, suggesting potential applicability to other environments, further research should build a larger dataset with more diverse environments and develop a model that is adaptable to different configurations.

\textbf{Consideration of virtual space.} In this work, when we collected data through simulation, we set the virtual space as infinitely empty and used the TAPF algorithm, which does not consider the virtual space. To ensure the generality of the virtual path, we simulated traversing a random 500 meter path 30 times and averaged them for a single placement. However, if objects or boundaries were present in the virtual space, the user's virtual path would follow patterns specific to the environment, potentially altering the optimal physical object placement for that virtual space. In such cases, the use of RDW algorithms that consider virtual space \cite{williams2021redirected, williams2021arc} could also affect the resets for the same physical space. Therefore, a potential direction for future work involves developing a model that predicts the number of resets by considering both virtual and physical spaces together, and experimenting with additional RDW algorithms.

\textbf{Interface considerations for practical application.}
While we have proposed an object placement interface with real-time feedback for RDW, there are further considerations to make the interface more usable in practice. First, the user should be able to easily select and position different objects to recreate an environment similar to their own space. The usability of such an interface can also be evaluated through user studies. Secondly, a system that recommends optimal arrangements could be beneficial. The optimal placement should not only minimize the number of resets but also take into account the context of the furniture and the relationships between the initial object placements.


\section{Conclusion}

We developed a Vision Transformer (ViT)-based model that predicts the number of resets RDW users will experience based on the placement of physical objects. We also presented an interactive interface that allows users to adjust object placement and receive real-time feedback on predicted resets. Our deep learning based approach can provide a rapid assessment of resets without the need for simulations or user studies.

While completely removing objects entirely from the physical space may be the simplest way to minimize resets, real-world environments are often constrained, making it impractical to freely move or remove objects. For example, heavy or plugged-in appliances, such as refrigerators, are not easy to relocate, and decorative items may be cumbersome to adjust. In some cases, there may simply be no alternative location for certain items. Our system offers a practical solution by helping users find an optimal arrangement of objects that works within these constraints. 

We believe that our approach can improve the RDW experience by giving users the ability to adjust their physical environment in a way that minimizes resetting while taking into account real-world constraints. Future research could extend the work by developing models that consider both physical and virtual spaces and incorporating larger datasets.

\begin{acks}
This work was supported by the National Research Foundation of Korea (No. RS-2024-00348094) and Korea Radio Promotion Association (No. RNIX20230200) grant funded by the Korea government (MSIT).
\end{acks}

\bibliographystyle{ACM-Reference-Format}
\bibliography{main}


\end{document}